# DATABASES OF PUBLICATIONS AND OBSERVATIONS - AS A PART OF THE CRIMEAN ASTRONOMICAL VIRTUAL OBSERVATORY

A. Shlyapnikov[1], N. Bondar'[1] and M. Gorbunov[1]

[1] *Stellar Physics Department, SRI "Crimean Astrophysical Observatory", 298409, Nauchnyj, Republic of Crimea; aas@crao.crimea.ua*

**Abstract.** The paper presents the basic principles of formation of a database (DB) with information about objects and their physical characteristics from observations carried out at the Crimean Astrophysical Observatory (CrAO) and published in "Izvestiya Krymskoi Astrofizicheskoi Observatorii" and other publications. The emphasis is placed on DBs that are not present in the most complete global library catalogs and data tables – VizieR (supported by the Strasbourg ADC). Separately, we consider the formation of a digital archive of observational data obtained at CrAO – as the interactive DB related to the DB of objects and publications. Examples of all the above DB as elements integrated into the Crimean Astronomical Virtual Observatory are presented in the paper. The operation with CrAO database is illustrated using tools of the International Virtual Observatory – Aladin, VOPlot, VOSpec jointly with VizieR DB and Simbad.

1. INTRODUCTION

The history of scientific studies in the Crimean Astrophysical Observatory accounts for more than a century as a whole. The amateur astronomy industrialist N.S. Maltsov donated his observatory to the Russian Academy of Sciences in 1908. In 1945 the department of Pulkovo Observatory in Simeiz was transformed into the independent organization. In the second half of the XX century CrAO became well-known in the world. At the beginning of the century the observatory operated with a small 0.12-m double astrograph, but as early as in 1925 the largest in the world at that time 1-meter telescope of the British firm Grubb-Parsons was already mounted in Simeiz. During the World War II the telescope was destroyed. In 1952 instead of the destroyed 1-m telescope the 1.22-m Zeiss reflector was mounted and put into regular operation at the new observatory (in Nauchnyj). The 2.6-meter reflecting telescope named after academician G.A. Shajn (ZTSh) was mounted in 1960. In 1966 the 22-meter radio telescope RT-22 was mounted in Katsiveli (near Simeiz). In 1974, after the reconstruction, the Tower Solar Telescope (TST-1) was put into operation. The unique gamma-ray telescope consisting of 48 mirrors with a total area of 54 sq. meters was brought into operation at CrAO in 1989. All the above instruments, at the time of their mounting, were the largest in Europe.

A significant contribution to the development of extra-atmospheric research was made at the Observatory after the start of the space age. Since 1959 fourteen instruments, designed and developed with participation of CrAO have been mounted on satellites and space stations. The most significant among these instruments were the orbital solar telescope (OST-1) mounted onboard Salut-4 space station in 1975 and ultraviolet telescope SPIKA with a primary mirror of 0.8 m in diameter onboard ASTRON space station that operated from 1983 to 1989.

At present, the observatory DB includes 32 observing instruments which were operated in conjunction with 55 devices that recorded radiation within the range from gamma rays with energies of $E > 10^{12}$ eV up to radio emission with a wavelength of 20 cm.

Such a considerable arsenal of observing equipment made it possible to solve a wide range of tasks, covering almost all the areas in the development of astronomy in the XX century. This was encouraged by the work of outstanding scientists – founders of scientific schools and prominent astrophysicists whose ideas laid the foundation for promising researches.

In the structure of CrAO databases DB "Staff" has a key connecting role between DBs "Publications" and "Observations". The above databases, along with databases "Projects", "Instruments", "Objects" and "Catalogs" constitute the basic part of the project "Crimean Astronomical Virtual Observatory – CrAVO".

More detailed information about the history of CrAO is to be found in the 5th edition of Volume 104 of "Izvestiya Krymskoi Astrofizicheskoi Observatorii", which presents materials of the conference dedicated to the 100th anniversary since the foundation of the observatory. CrAVO project has been described by one of the authors (Shlyapnikov 2007).

2. DB «PUBLICATIONS» (DBP)

The database is formed from publications of researchers based on observations carried out at CrAO and/or in other observatories. It includes studies implemented in other organizations where our researchers were co-authors. The first volume of "Izvestiya Krymskoi Astrofizicheskoi Observatorii" was published in 1947. The DBP involves 100 volumes, since the subsequent volumes are available in interactive mode and search for information in them is less difficult compared to the printed version. Before the release of the first volume of "Izvestiya" papers of researchers had been published in various journals, but then a dominant number of papers to the end of the XX century were published in "Izvestiya Krymskoi Astrofizicheskoi Observatorii".

The motivation for creating DBP was the analysis of how "Izvestiya Krymskoi Astrofizicheskoi Observatorii" are represented in the most significant world database of astronomical publications – digital library portal of the Smithsonian Astrophysical Observatory for researchers in astronomy and physics supported by grants from NASA (SAO/NASA Astrophysics Data System - ADS) (Kurtz M.J. et al. 2000).

By the mid 2013 (time of the latest analysis) the ADS search system had provided a gateway to information on 1283 papers published in "Izvestiya Krymskoi Astrofizicheskoi Observatorii". Note that over six years (Shlyapnikov 2007), the number of references to "Izvestiya" has enlarged up to 406 publications; however, it is only 57% of the entire list of publications. The main suppliers of information for ADS are: program of scientific and technical information NASA (609 publications); Center for Astronomy of Heidelberg University (Germany) (491 publications), and Strasbourg Astronomical Data Center (France) (136 publications). The interest of various organizations to this journal is associated with information contained in papers, including observational results.

However, the main problem is still unavailability to access original publications and, in some cases, their abstracts. So, to improve this situation, we have undertaken a work on the digitization of papers in "Izvestiya of KrAO".

In the paper by Bondar' et al. (2013) we discussed in detail the problem of creating a digital version of "Izvestiya". The basic and most time-consuming work in creating DBP is compiling a subject index (indexing of content – IC) for published articles. IC is not a part of the structure of "Izvestiya of KrAO", what makes it impossible to conjugate databases "Publications" and "Objects". We have started work on compiling IC. The example of its brief format is shown in Table 1. Here is a reference to "Izvestiya" and object/objects described in this paper. The extended format of IC is supplemented with information about the instrument, used in observations, characteristics of data (photometric, spectral, polarimetric, etc.) and time of observation.

Table 1. Example of indexing journal "Izvestiya of KrAO".

| IzKry | NAME | IzKry | NAME | IzKry | NAME |
|---|---|---|---|---|---|
| 1966..66.139 | MetS | 1976..55..85 | HD 224085 | 1978..58..51 | Cyg X-3 |
| 1975..53..29 | GRS | 1976..55..94 | BY Dra | 1978..58..56 | V 599 Aql |
| 1975..53..59 | Cyg X-3 | 1976..55.100 | EM Cep | 1978..58.104 | NGC 4151 |
| 1975..53..59 | Cas GAM-1 | 1976..55.112 | OAs* | 1979..59.104 | Jupiter |
| 1975..53.150 | Fl* | 1976..55.112 | YCl* | 1979..59.133 | CI Cyg |
| 1975..53.154 | RD* | 1976..55.127 | Sp.* | 1979..59.143 | AN And |
| 1975..53.165 | Fl* | 1976..55.157 | Cyg X-3 | 1979..59.182 | Mrk 279 |

In the process of developing a digital version of "Izvestiya of KrAO" abstracts of papers, GIF and PDF formats of publications and contained data are prepared for inclusion into the Vizier database (Ochsenbein F. et al. 2000) and SAO/NASA ADS.

Database "Publications", supplemented with IC, includes hyperlinks to ADS database and SIMBAD (Wenger M. et al. 2000) and provides interaction with database "Observations" of CrAVO.

3. DB «OBSERVATIONS» (DBO)

The developing of digital databases of observations carried out with different instruments has been started since the mid-90s of the last century at CrAO. The process of forming database of photographic archives, creation of their digital version, and possibility to use it for solving astrophysical tasks were described in detail earlier in a series of our publications (Bondar' 1999, Bondar' 2002, Bondar' et al. 2005, Bondar' & Shlyapnikov 2006, Bondar' & Shlyapnikov 2009). Below we present some results of this work. DBO according to specifics of the obtained material is to be conditionally divided into 3 archives: direct images of the sky; negatives obtained with objective prisms; spectral photographic observations derived at 1-m, 1.22-m and 2.6-m telescopes.

The largest collection of direct images (~ 1500) was obtained in the process of implementation of "G.A. Shajn's Plan" to study structure of the Galaxy over the period 1947–1965 (Pronik 1998) and as a result of work on the "Crimean Review of Minor Planets" (~ 10000) in the period 1963–1999 (Chernykh 1992). The distribution of the sky negatives obtained from observations on these projects is shown in Fig. 1. DBO are prepared in a variety of formats, including recommended for work with applications of the International Virtual Observatory (IVO). Fig. 1 illustrates DBO using the IVO application Aladin (Bonnarel F. et al. 2000).

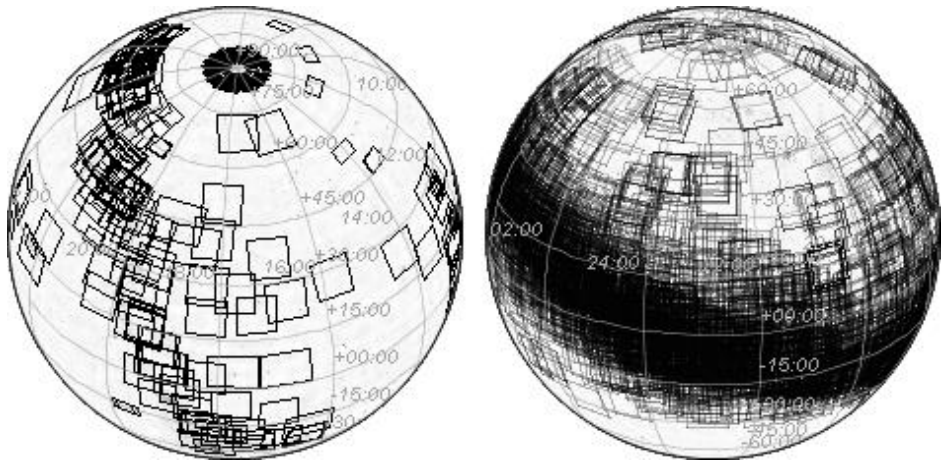

Fig. 1. Negatives obtained on "G.A.Shajn's Plan" (left panel) and "Crimean Review of Minor Planets" (right panel) projects distributed over the celestial spheres

An example of the IVO application Aladin for working with DBO is shown in Fig. 2. Here, in the central part of interface – preview image of one of the negatives, at the bottom – database of observations with hyperlinks to the digital version of small copies intended for previewing and assessing quality of photographic plates. Access to the local archive with DBO and preview images from the collection of negatives on "G.A.Shajn's Plan" is organized via the portal of Ukrainian Virtual Observatory (http://ukr-vo.org/digarchives/ index.php?b5&1) (Vavilova et al. 2012a, Vavilova et al. 2014), or via CrAO server - http://www.crao.crimea.ua/~aas/PROJECTs.

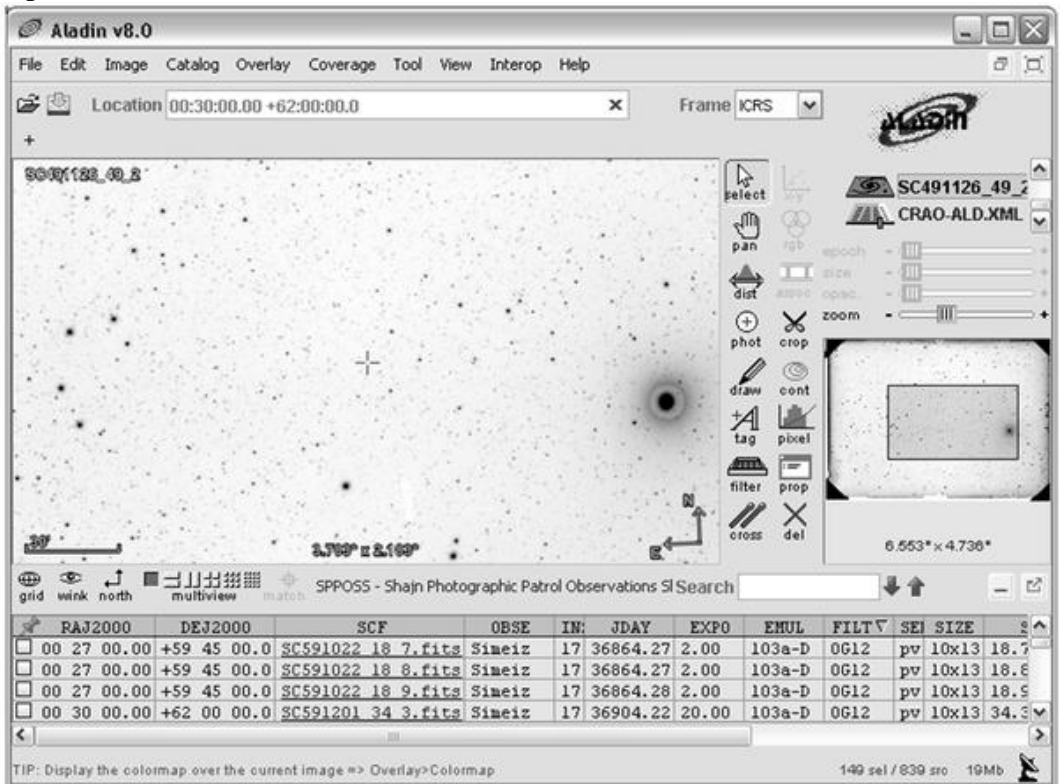

Fig. 2. Example of using IVO application Aladin for working with DBO, obtained under the "G.A.Shajn's Plan" project.

Information about the described above archives is placed at WFPA (http://www.skyarchive.org/), and observational data – in WFPDB and available with a key «CRI» at the address: http://draco.skyarchive.org/search/, or via the VizieR database in catalog VI/90 (Tsvetkov et al. 1997). Prospects for scientific researches with DBO and archive of direct images are described in detail in papers by Bondar' & Shlyapnikov (2009) and Vavilova et al. (2012b).

Along with direct images there is a number of negatives with spectral observations in the archive of CrAO. Over the period from 1929 to 1992, about 15000 spectra have been taken with various instruments at CrAO. Among them: 1340 negatives obtained from 1929 to 1941 at the 1-m telescope in Simeiz; more than 500 plates acquired with objective prisms from 1929 to 1965 at the 0.17-m and 0.4-m astrographs, including negatives from "G.A.Shajn's Plan» (Pronik 1998); 5570 plates of different spectral resolution in different ranges obtained from 1953 to 1990 at the 1.22-m Zeiss reflector (Shlyapnikov 2013); 2900 spectra – at different spectrographs and with different spectral resolution in 1963–1987 at the 2.6-m telescope, named after academician G.A. Shajn (ZTSh), and 3450 spectra – at the spectrograph with image converter (ZTSh, Nasmyth focus) in 1982-1992 (Polosukhina et al. 1997). Examples of digital versions of spectral archives are described in Gorbunov & Shlyapnikov (2013) and Pakuliak et al. (2014).

In the process of developing a digital version of archives of spectral observations, special attention should be paid to the preparation of data in a format compatible with the used means of IVO and their representation in the interactive applications Aladin, VOSpec, Specview. The photographic spectral archives are divided into the collection of negatives obtained at three astrographs with objective prisms and obtained at three large telescopes (Fig. 3).

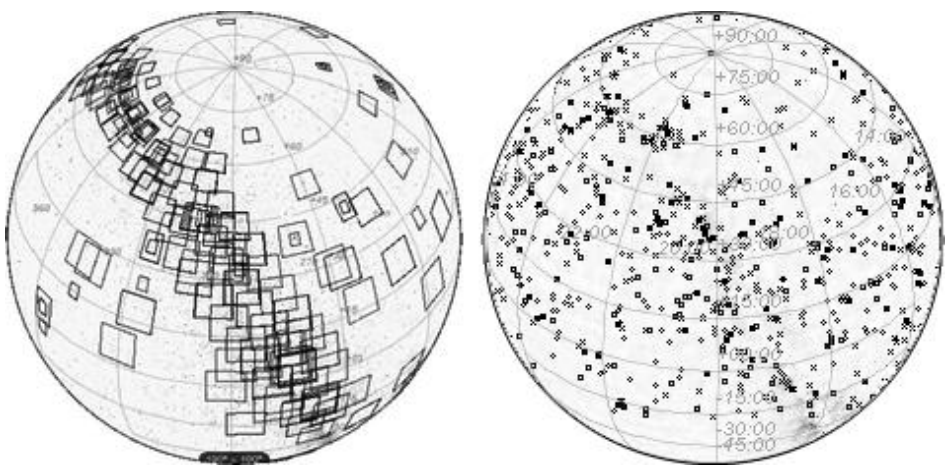

Fig. 3. Spectral plates obtained at astrographs with objective prisms (left panel) and 1-m (cross), 1.22-m (rhombus) and 2.6-m (square) telescopes (right panel) distributed over the celestial spheres

Below (Fig. 4) there are examples of presentation of the spectrum alf And in program Specview (obtained at the 1-m telescope in Simeiz on 17 November 1929) with identification of spectral lines (top panel) and in the VOSpec application of infrared spectroscopic observations of eps Aur (acquired at the 1.22-m telescope with image converter FTC 1A) compared with data for an object taken from IVO databases (bottom panel).

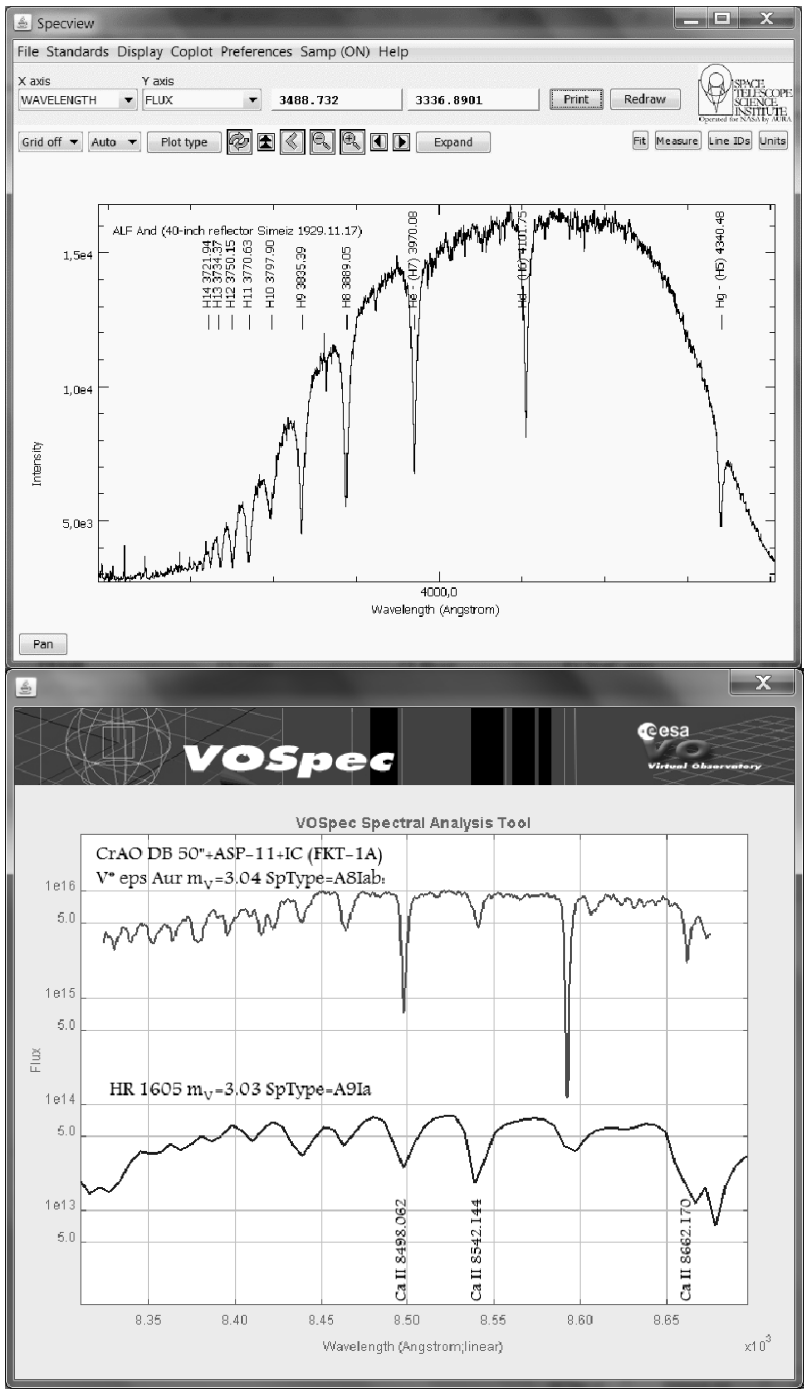

Fig. 4. Example of presenting spectral observations from CrAO DBO collection using IVO tools.

## 3. CONCLUSIONS

As a result of this work we outlined the main goal of further development of databases "Publications" and "Observations", their integration into the CrAVO, in conjunction with DBs "Staff", "Projects", "Instruments", "Objects" and "Catalogs".

The current state of the database "Publications". DBP contains information on nearly 1500 papers written with participation of CrAO researchers. These publications include data for more than 20000 objects involved in database "Observations." DBO partially integrated into the VizieR database as well as archive of the CrAVO are available via interactive applications IVO.

ACKNOWLEDGMENTS. Authors thank L.P. Metik, N.I. Yavorskaya and N.V. Elizarova for a great work on inputting data from journals of observations into a computer. It was a great help in preparing database "Observations". Authors express their sincere gratitude to R.E. Gerchberg for constant attention to the work on developing CrAVO and valuable comments on the text of the paper. Many thanks to Ya.V. Poklad for editing the English version of this paper.

In our research we made use of the SIMBAD database, VizieR catalogues and "Aladin sky atlas" developed and supported at CDS, Strasbourg Observatory, France and SAO/NASA Astrophysics Data System, USA. Authors are thankful to all who created this.

N.I. Bondar' grateful for partial support of this work of the Russian Foundation for Basic Research (project 15–02–06271).